\documentclass[utf8]{frontiersSCNS}

\usepackage{url,hyperref,microtype,subcaption}
\usepackage{ulem}
\usepackage{multirow}

\newcommand{\MBH}{$M_{\rm BH}$}

\newcommand{\Lbol}{$L_{\rm bol}$}
\newcommand{\Avneb}{$A_{V}^{\rm neb}$}

\newcommand{\hii}{\ifmmode \rm{H}\,\textsc{ii} \else H~{\sc ii}\fi}
\newcommand{\Ha}{\ifmmode {\rm H}\alpha \else H$\alpha$\fi}
\newcommand{\Hb}{\ifmmode {\rm H}\beta \else H$\beta$\fi}
\newcommand{\oiii}{\ifmmode [\rm{O}\,\textsc{iii}] \else [O~{\sc iii}]\fi}
\newcommand{\oii}{\ifmmode [\rm{O}\,\textsc{ii}] \else [O~{\sc ii}]\fi}
\newcommand{\oi}{\ifmmode [\rm{O}\,\textsc{i}] \else [O~{\sc i}]\fi}
\newcommand{\nii}{\ifmmode [\rm{N}\,\textsc{ii}] \else [N~{\sc ii}]\fi}
\newcommand{\sii}{\ifmmode [\rm{S}\,\textsc{ii}] \else [S~{\sc ii}]\fi}

\def\keyFont{\fontsize{8}{11}\helveticabold }
\def\firstAuthorLast{Kozie\l -Wierzbowska {et~al.}} 
\def\Authors{Dorota Kozie\l -Wierzbowska\,$^{1,*}$, Gra\.zyna Stasi\'nska\,$^{2}$, Natalia Vale Asari\,$^{3}$, Marek Sikora\,$^{4}$, Elisa Goettems\,$^{3}$ and Anna W\'ojtowicz\,$^{1}$ }
\def\Address{$^{1}$Astronomical Observatory, Jagiellonian University, ul. Orla 171, PL-30244 Krakow, Poland \\
$^{2}$LUTH, Observatoire de Paris, CNRS, Universit\'e Paris Diderot; Place Jules Janssen 92190 Meudon, France\\
$^{3}$Departamento de F\'{\i}sica--CFM, Universidade Federal de Santa Catarina, C.P.\ 476, 88040-900, Florian\'opolis, SC, Brazil\\
$^{4}$Nicolaus Copernicus Astronomical Center, Bartycka 18, 00-716 Warsaw, Poland  }
\def\corrAuthor{Dorota Kozie\l -Wierzbowska}

\def\corrEmail{dorota.koziel@uj.edu.pl}

\begin{document}

\onecolumn
\firstpage{1}

\title[Pair-matching of radio-loud and radio-quiet AGNs]{Pair-matching of radio-loud and radio-quiet AGNs} 

\author[\firstAuthorLast ]{\Authors} 
\address{} 
\correspondance{} 

\extraAuth{}

\maketitle

\begin{abstract}
\tiny
 \keyFont{ \section{Keywords:} active galaxies, radio galaxies, jetted and non-jetted AGNs, radio-loudness, galaxy morphology}

 Active galactic nuclei (AGNs) are known to cover an extremely broad range of radio luminosities and the spread of their radio-loudness is very large at any value of the Eddington ratio. This implies very diverse jet production efficiencies which can result from the spread of the black hole spins and magnetic fluxes. Magnetic fluxes can be developed stochastically in the innermost zones of accretion discs, or can be advected to the central regions prior to the AGN phase. In the latter case there could be systematic differences between the properties of galaxies hosting radio-loud (RL) and radio-quiet (RQ) AGNs. In the former case the differences should be negligible for objects having the same Eddington ratio.  To study the problem we decided to conduct a comparison study of host galaxy properties of RL and RQ AGNs. In this study we selected type II AGNs from SDSS spectroscopic catalogues. Our RL AGN sample consists of the AGNs appearing in the \citet{Best.Heckman.2012a} catalogue of radio galaxies. To compare RL and RQ galaxies that have the same AGN parameters we matched the galaxies in black hole mass, Eddington ratio and redshift. We compared several properties of the host galaxies in these two groups of objects like galaxy mass, colour, concentration index, line widths, morphological type and interaction signatures. We found that in the studied group RL AGNs are preferentially hosted by elliptical galaxies while RQ ones are hosted by galaxies of later type. We also found that the fraction of interacting galaxies is the same in both groups of AGNs. These results suggest that the magnetic flux in RL AGNs is advected to the nucleus prior to the AGN phase.

\end{abstract}

\section{Introduction}
Active galactic nuclei (AGNs) are powered by accretion on a supermassive black hole (BH). Although the first discovered AGNs were radio-loud (RL), most of AGNs are radio-quiet (RQ). The radio-loudness of RL AGNs, defined as the ratio of the radio flux to the optical flux \citep{Kellermann.etal.1989a}, covers several orders of magnitude \citep[e.g.][]{Sikora.Stawarz.Lasota.2007a,Lal.Ho.2010a} which implies very diverse jet production efficiencies. For jets powered by rotating BHs \citep{1977MNRAS.179..433B}, the efficiency of the jet production can be related to the spread of the BH spins and the amount of magnetic fluxes accumulated in the central AGNs.

Magnetic fluxes can be developed stochastically in the innermost zones of accretion discs \citep{2014ApJ...782L..18B}, or can be advected to the central regions of a galaxy prior to the AGN phase \citep{Sikora.etal.2013a,Sikora.Begelman.2013a}. In the latter case there could be systematic differences between the properties of galaxies hosting radio-loud and radio-quiet AGNs. Our aim is to compare the properties of the host galaxies of radio-loud and radio-quiet AGNs to distinguish which of these two scenarios is more probable. If we find that RQ and RL are hosted by different galaxies, we could discard the scenario where radio jets are stochastic.

Radio-loud (jetted) and radio-quiet (non-jetted) AGNs have already been studied extensively and some systematic differences were found between these two groups of objects:
\begin{itemize}
\item The most radio-loud AGNs are found in galaxies with black hole masses $\ge 10^8\, M_{\odot}$ \citep{Laor2000a,Dunlop.etal.2003a,Floyd.etal.2004a,McLure.Jarvis.2004a,Best.etal.2005a}.
\item The fraction of radio-loud objects and radio loudness increases with decreasing Eddington ratio, $\lambda=$\Lbol$/L_{\rm Edd}$  \citep[e.g.][]{Kratzer.Richards.2015a,Terashima.Wilson.2003a}, but there is a large scatter in radio loudness for AGNs with similar Eddington ratio.
\item The fraction of galaxies with disturbed morphology is larger in RL than in RQ AGNs \citep{Bessiere.etal.2012a,Chiaberge.etal.2015a}.
\item RL AGNs are located in denser environments \citep[e.g.][]{Mandelbaum.etal.2009a,RamosAlmeida.etal.2013a}.  
\end{itemize}

However, the differences listed above concern entire families of objects, but not objects that have the same accretion properties.  Moreover, in catalogues of AGNs there are many objects with properties characteristic of RL objects like very massive black holes, low Eddington ratios, and  disturbed morphologies, but they are radio-quiet. Therefore, we ask the question why is the efficiency of the jet production very different among otherwise similar objects?

In our project we concentrate on Type 2 (i.e. obscured) objects, to be able to study the properties of the host galaxies, with Eddington ratios $\lambda\,\geq$ 0.003, and we seek the differences between radio-loud and radio-quiet AGNs to check if there are any generic differences between the host galaxies of these two groups of objects that can indicate different evolution histories of jetted and non-jetted AGNs.

\section{Methods}

\subsection{Selection of the samples}

The samples of radio-loud and radio-quiet galaxies were selected from the SDSS DR7 spectroscopic catalogues \citep{Abazajian.etal.2009a}. Only galaxies with S/N $\geq$ 10 \AA$^{-1}$ in the region around $\lambda_0$ = 4020 \AA\ were chosen. We also applied the redshift cuts, low velocity dispersion limit and S/N limits in emission line fluxes as in  \citet{KozielWierzbowska.etal.2017a}. Galaxies with faulty pixels in the area of emission lines were eliminated. After this step, AGNs were selected based on the BPT diagram and the \citet{Kewley.etal.2001a} line. 
Using the WHAN diagram \citep{CidFernandes.etal.2011a}, we removed those galaxies in which the emission lines could be produced by hot, low-mass, evolved stars (the retired galaxies  defined in \citealp{Stasinska.etal.2008a}).
This procedure selected 19883 optical AGNs.

Among  this sample of optical AGNs, we searched those which belong to the \citet[][BH12]{Best.Heckman.2012a} catalogue of radio galaxies and whose radio emission is considered by them to be produced by an AGN. The BH12 catalogue flux limit is 5 mJy.

After limiting ourselves to AGNs with Eddington ratio $\lambda \geq$ 0.003\footnote{This step ensures us that we calculate the bolometric luminosity consistently and correctly for all studied sources.}, i.e. focusing mostly on sources with radiatively efficient accretion, we obtained our RL AGN sample of 376 objects, and our RQ AGN sample of 10918 objects. 

The host galaxy stellar masses, velocity dispersions (used to calculate \MBH),  nebular extinction,  emission line fluxes and equivalent widths, and the Eddington ratios, $\lambda$, where obtained from the SDSS data after applying the STARLIGHT \citep{CidFernandes.etal.2005a} spectral model-fitting.

\begin{figure}[h!]
\begin{center}
\includegraphics[width=12cm]{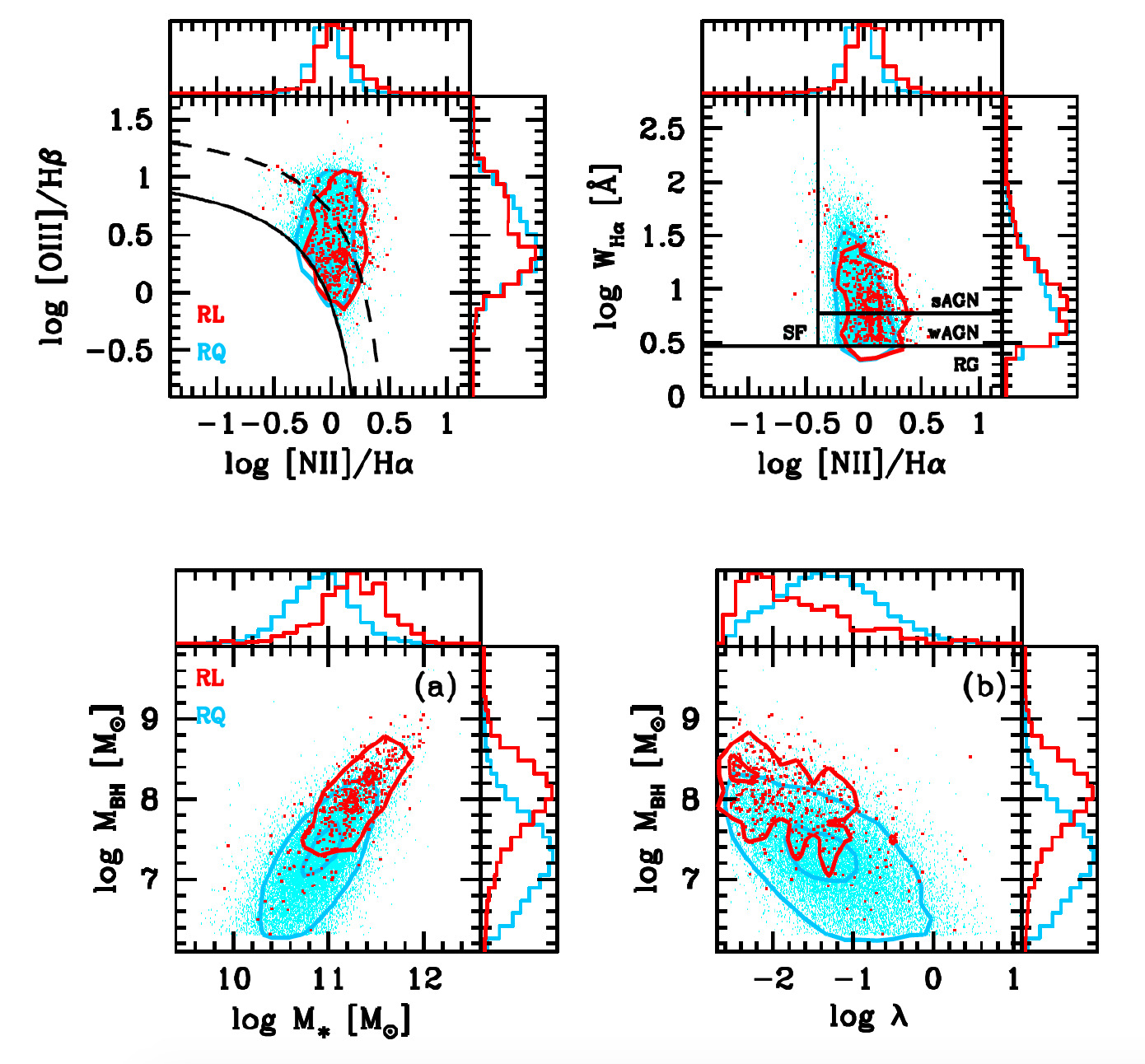}
\end{center}
\caption{Top: BPT and WHAN diagrams for RL and RQ AGNs (red and blue points, respectively). The solid black line in the BPT diagram is the \citet{Kewley.etal.2001a} line, while the dashed line separates galaxies with line emission dominated by the AGN. In the WHAN diagram the solid lines separate SF galaxies, strong AGNs, weak AGNs and retired galaxies \citep{CidFernandes.etal.2011a}. Bottom: $M_{BH}$ versus stellar mass and $M_{BH}$ versus Eddington ratio. The histograms of plotted parameters for RL and RQ galaxies are also shown. Updated version published in \citet{KozielWierzbowska.etal.2017a}. }\label{fig:1}
\end{figure}
 
The upper panels of Fig. \ref{fig:1} show the BPT \citep{Baldwin.Phillips.Terlevich.1981a} and the WHAN \citep{CidFernandes.etal.2011a} diagrams for our RL and RQ samples. RL AGNs are shown in red and RQ AGNs in blue. On both axes the normalized histograms are presented. The black curve is the \citet{Kewley.etal.2001a} line. The location of red and blue points is similar which shows that objects from both samples have similar ionization properties. 
The bottom panel in Fig. \ref{fig:1} shows $M_{BH}$ as a function of the galaxy stellar mass $M_{*}$, and of the Eddington ratio, $\lambda$. In these diagrams RL and RQ galaxies are shifted relative to each other. On average, RL AGNs have higher black hole masses and higher stellar masses compared to RQ galaxies. RL objects have also lower $\lambda$. These observations are compatible with trends found by previous authors (see Introduction). These two panels clearly show that in order to compare RL and RQ AGNs it is necessary to match them in black hole mass and Eddington ratio. 

\subsection{Pair-matching of RL and RQ galaxies}

To compare RL and RQ galaxies with the same AGN parameters, we applied a pair-matching technique. Galaxies were matched in redshift, black hole mass ($M_{BH}$) and Eddigton ratio ($\lambda$). 
In practice for each RL object we selected all RQ objects for which: $|\Delta z| \le 0.01$, $|\Delta \log \lambda| \le 0.09$, and $|\Delta \log M_{\rm BH}| \le 0.1206$. We computed the distance as $d^2_{\rm match} = \sum \Delta^2$. For each RL AGN three RQ galaxies with the smallest $d$ were included into the matched RQ sample (mRQ sample). 

We defined the radio loudness parameter by ${\mathcal R} \equiv L_{\rm 1.4} [W\, Hz^{-1}] / L_{\Ha} [ L_{\odot}]$, where $L_{\rm 1.4}$ is the radio luminosity at 1.4 GHz.
The classical criterion for radio-loud AGN, ${\mathcal R}^{(K)} > 10$ \citep{Kellermann.etal.1989a}, translates into $\log {\mathcal R} > 15.8$. From the matched RQ sample we excluded galaxies that were undetected in radio and for which the radio-loudness (estimated from the flux limit od 5 mJy in the BH12 catalogue) is larger than 10 ($\log {\mathcal R} > 15.8$). 

\section{Results}

\begin{figure}[h!]
\begin{center}
\includegraphics[]{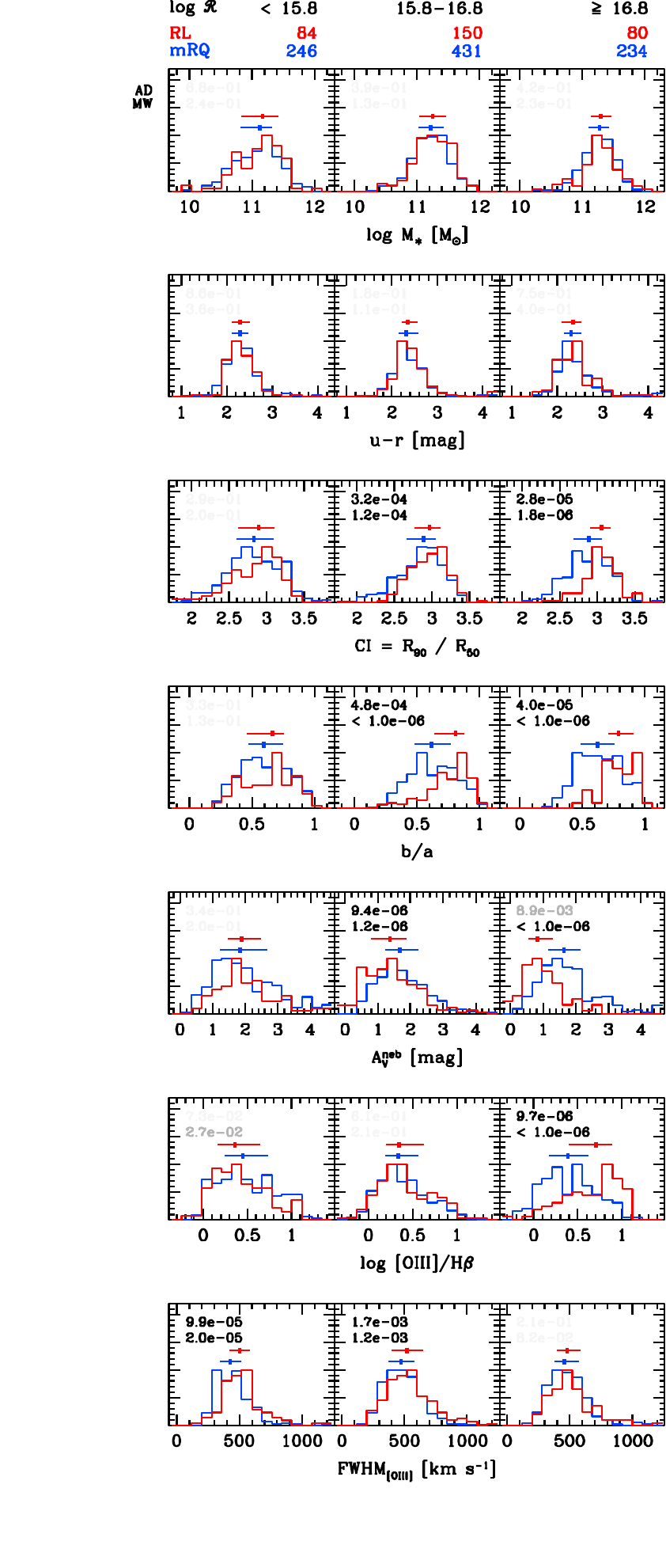}
\end{center}
\caption{Histograms of selected parameters of RL (red) and mRQ (blue) objects. In each panel we show the results of the Anderson-Darling and Mann-Whitney tests in terms of probabilities p that the samples could be drawn from the same population. Results with $p<0.003$ are written in black, those with $p< 0.05$ are written in dark grey. Blue and red points at the top of each histogram indicate the median values and the horizontal lines indicate the quartiles. Published in \citet{KozielWierzbowska.etal.2017a}}\label{fig:2}
\end{figure}

We present our results in the form of histograms. Fig. \ref{fig:2} shows histograms of selected parameters of RL (red) and RQ (blue) galaxies. Results for RL AGNs and their matched RQ galaxies are presented separately for three bins in the value of the radio-loudness ${\mathcal R}$ of the parent RL object. ${\mathcal R}$ bins were chosen to represent classically radio-quiet (although jetted, $\log {\mathcal R} < 15.8 $) objects, intermediate radio-loud objects ($15.8 < \log {\mathcal R} < 16.8$), and the most radio-loud ones ($\log {\mathcal R} > 16.8$). In each bin, the tick mark with horizontal line shown on the top of the histograms represents the median value and the quartiles. We used two statistical tests to study the difference between RL and mRQ samples. These are the Anderson-Darling (AD) and Mann-Whitney (MW) tests. These tests give the probability that both samples are drawn from the same population.

Among the studied parameters, the stellar galaxy masses, $M_{*}$, and the colors, {\it u-r}, have very similar distributions for RL and RQ AGNs. In the case of galaxy mass this result is expected since we matched our objects in $M_{BH}$ which correlates with $M_{*}$. The concentration index, CI, and the galaxy axes ratio, $b/a$, the two parameters associated with galaxy shape and morphology, show significant differences. RL galaxies tend to have larger concentration index, and larger axes ratios in the two highest ${\mathcal R}$ bins. The lower values of CI and $b/a$ ratio in the matched RQ sample indicate more disky galaxy morphology. 

In the next two panels we present a comparison of the nebular extinction, \Avneb,\ and of the \oiii\ to \Hb\ line fluxes ratio. The values of \Avneb\ decrease with increasing ${\mathcal R}$. \oiii$/$\Hb\ differs significantly in the bin of the highest ${\mathcal R}$. These results may suggest some contribution from the \hii\ regions to the line emission. The last panel shows histograms of \oiii\ line widths. We see that the line widths in RL galaxies tend to be larger then in RQ AGNs in the lowest ${\mathcal R}$ bins. We speculate that this can result from having at low R less collimated jets. Forming wider outflows such jets may affect the motion of the gas clouds in the narrow line region, and hence broaden the emission line profiles. 

To eliminate the contribution of \hii\ regions from our studies we decided to use ``cleaned'' samples. In these samples only galaxies that lie above the dashed curve shown in BPT diagram from Fig. \ref{fig:1} are included. In these galaxies the contribution from \hii\ regions can be neglected. The results for the cleaned samples \citep[see][]{KozielWierzbowska.etal.2017a} confirm our findings for the whole RL and mRQ samples concerning morphological properties. 

\begin{figure}[h!]
\begin{center}
\includegraphics[width=16cm]{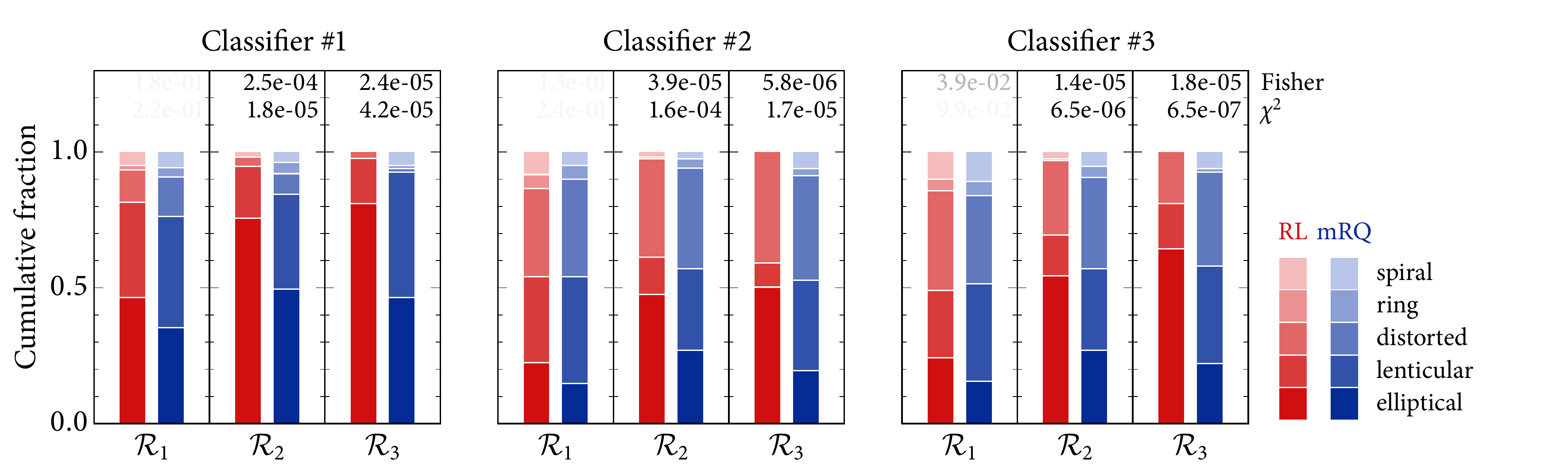}
\end{center}
\caption{Cumulative fraction of RL (shades of red) and RQ (shades of blue) classified as elliptical, lenticular, distorted, ring, or spiral galaxies. $R_{1}$ corresponds to $\log {\mathcal R} < 15.8 $, $R_{2}$ to $15.8 < \log {\mathcal R} < 16.8$, and $R_{3}$ to $\log {\mathcal R} > 16.8$. Results are shown for three classifiers separately.}\label{fig:3}
\end{figure}

From the comparison of CI and of the $b/a$ ratio we see that there is a difference in morphology between RL and mRQ galaxies. To confirm this result we decided to perform a morphological classification of all RL galaxies and the closest match from the RQ galaxies. Classifiers looked at the color SDSS images of these objects and attributed to each galaxy a morphological type. The results for the three classifiers are shown in Fig. \ref{fig:3}. The panels in this figure show in different shades of red (blue) the fraction of RL (mRQ) galaxies classified as elliptical, lenticular, distorted, ring, or spiral galaxies. As before, results are presented separately for three ${\mathcal R}$ bins. 
As we can see, the fraction of ellipticals among RL objects in all three radio-loudness bins is larger than among the matched RQ galaxies. Note that low-CI galaxies are not spirals, but lenticular or distorted galaxies. This result was confirmed using Galaxy ZOO data. 

\begin{figure}[h!]
\begin{center}
\includegraphics[width=16cm]{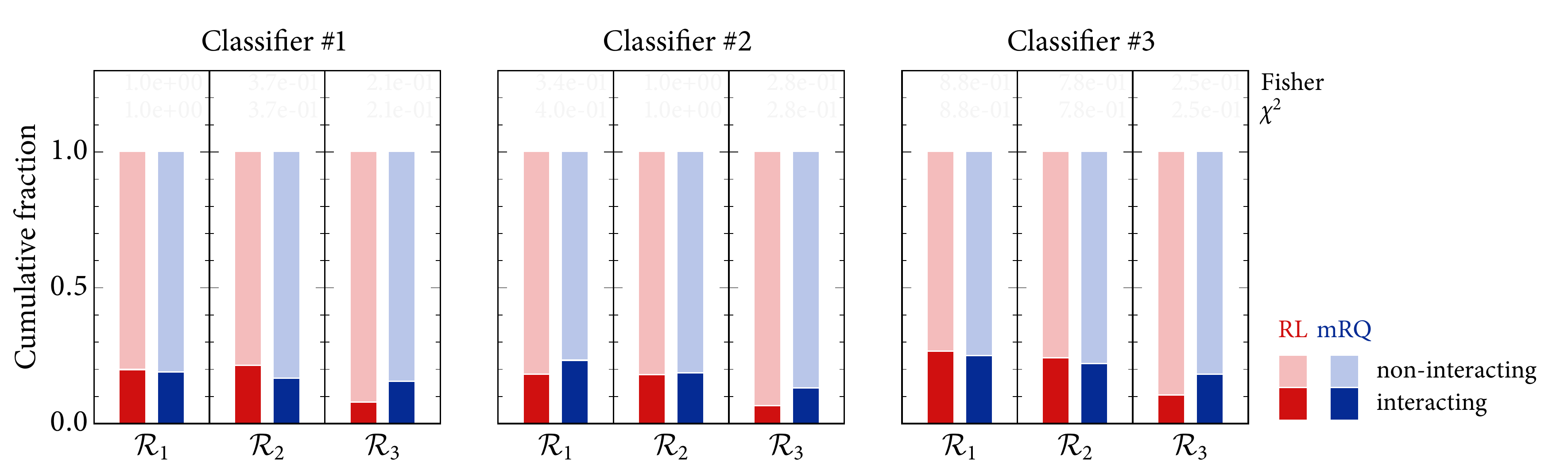}
\end{center}
\caption{Cumulative fraction of RL AGNs in three ${\mathcal R}$ bins, and their closest mRQ galaxies showing signs of the galaxy interaction. The layout is the same as for Fig. \ref{fig:3}.}\label{fig:4}
\end{figure}

The classification scheme also included information about galaxy interactions like major or minor mergers, tail, or suspected interaction. Fig. \ref{fig:4} shows the results concerning  the fraction of galaxies with interaction signatures considered as certain. Here, as interacting, we consider major or minor mergers and galaxies with a tail. In Fig. \ref{fig:4} we do not see any systematic difference between RL and RQ AGNs suggesting that galaxy interaction has no special effect on the radio activity of the AGN. Our result is in contradiction with \citet{Chiaberge.etal.2015a} who found that all radio-loud galaxies in their sample are mergers, however their sample consisted of FRII radio galaxies at redshifts larger than 1, while in our sample we have only local AGNs mostly with compact morphologies. 

\section{Summary}

The results presented here clearly show that not only Eddington ratio and black hole mass determine the jet production efficiency. In the description of the jet production the accumulation of the magnetic flux in the AGN center and the BH spin have also to be taken into account. The fact that we find a higher fraction of ellipticals among RL galaxies than among the matched RQ ones suggest that elliptical galaxies have already a sufficient amount of magnetic flux accumulated in the center to produce and collimate radio jets in the \citealp{1977MNRAS.179..433B} scenario. This is in agreement with findings of \citet{Sikora.etal.2013a} and \citet{Sikora.Begelman.2013a} on the RL AGN pre-phase. 

\section*{ACKNOWLEDGMENTS}

This work was carried out within the framework of the Polish National Science Centre grant UMO-2013/09/B/ST9/00026.  Gra\.zyna Stasi\'nska was partially supported by the National Research Centre, Poland, DEC-2013/08/M/ST9/00664, within the framework of the HECOLS International Associated Laboratory. GS and NVA acknowledge the support from the CAPES CSF--PVE project 88881.068116/2014-01. The Sloan Digital Sky Survey is a joint project of The University of Chicago, Fermilab, the Institute for Advanced Study, the Japan Participation Group, the Johns Hopkins University, the Los Alamos National Laboratory, the Max-Planck-Institute for Astronomy, the Max-Planck-Institute for Astrophysics, New Mexico State University, Princeton University, the United States Naval Observatory, and the University of Washington.  Funding for the project has been provided by the Alfred P. Sloan Foundation, the Participating Institutions, the National Aeronautics and Space Administration, the National Science Foundation, the U.S. Department of Energy, the Japanese Monbukagakusho, and the Max Planck Society.

\bibliographystyle{frontiersinSCNS_ENG_HUMS} 
\bibliography{references}

\end{document}